%% file: paper.tex
\documentclass[a4paper]{article}

\usepackage{INTERSPEECH2020}

\title{Training Keyword Spotting Models on Non-IID Data with Federated Learning}
\name{Andrew Hard, Kurt Partridge, Cameron Nguyen, Niranjan Subrahmanya,
      Aishanee Shah, Pai Zhu, Ignacio Lopez Moreno, Rajiv Mathews}
\address{
  Google LLC,\\
  Mountain View, CA, U.S.A.}
\email{harda@google.com}

\begin{document}

\maketitle
\begin{abstract}
  We demonstrate that a production-quality keyword-spotting model can be trained
  on-device using federated learning and achieve comparable false accept and
  false reject rates to a centrally-trained model. To overcome the algorithmic
  constraints associated with fitting on-device data (which are inherently
  non-independent and identically distributed), we conduct thorough empirical
  studies of optimization algorithms and hyperparameter configurations using
  large-scale federated simulations. To overcome resource constraints, we
  replace memory-intensive MTR data augmentation with SpecAugment, which reduces
  the false reject rate by 56\%. Finally, to label examples (given the zero
  visibility into on-device data), we explore teacher-student training.
\end{abstract}
\noindent\textbf{Index Terms}: federated learning, on-device learning, keyword
spotting, wake word detection, non-iid data, data augmentation

\input{introduction}
\input{model}
\input{optimization}
\input{data}
\input{experiments}
\input{conclusions}
\input{acknowledgements}

\bibliographystyle{IEEEtran}

\bibliography{mybib}

\end{document}

%% file: introduction.tex
\section{Introduction}
\label{sec:introduction}

Keyword spotting has become an essential access point for virtual assistants.
Vocalized keywords such as \textit{Alexa}, \textit{Hey Google}, or
\textit{Hey Siri} can be used to initiate search queries and issue commands to
mobile phones and smart speakers. The underlying algorithms must process
streaming audio---the majority of which must be ignored---and trigger quickly
and reliably when needed.

Neural networks have achieved state-of-the-art performance in automatic speech
recognition tasks~\cite{dnn_asr, dnn_lvsr, e2e_sr}. Applications of neural
networks to keyword spotting have also been explored, particularly within the
contexts of quality improvement and latency reduction for low-resource
environments~\cite{small_fp_kws, mtl_wce_kws, hey_siri, comp_td_nn_kws, kws_smp,
cascade_kws} and end-to-end model training~\cite{e2e_kws}.

\textit{Federated Learning} (FL)~\cite{fedlearn} is a decentralized
computation paradigm that can be used to train neural networks directly
on-device. In FL, all model updates shared by devices with the server are
\textit{ephemeral} (only stored temporarily in server memory), \textit{focused}
(only relevant to a specific training task), and \textit{aggregated} (only
processed collectively with updates from other devices across the population).
In conjunction with techniques such as differential privacy~\cite{dp, dl_dp} and
secure aggregation~\cite{secagg}, FL can integrate strong anonymity and privacy
guarantees into the neural network training process.

Federated learning provides a path to train keyword models at the edge, on real
user data, as opposed to proxy data. In contrast, centrally-trained
keyword-spotting models use proxy data, since false accepts (in which the
keyword-spotter accidentally triggers) are not logged.

Multiple production models have been trained with federated learning, including
next-word prediction~\cite{flnwp}, emoji prediction~\cite{fl_emoji}, n-gram
language models~\cite{fl_ngram}, and query suggestions~\cite{fl_c2q} for mobile
keyboards. Many of these models achieve better performance as a result of the
additional signals and unbiased data available on-device. Recently, the
feasibility of training keyword-spotting algorithms with FL has been explored
with smaller datasets~\cite{fed_kws}.

On-device training comes with challenges, including the fact that the quantities
and characteristics of training examples vary considerably from device to
device. Centrally-trained models benefit from the ability to sample data in a
controlled, independent and identically distributed (IID) manner, resulting in
gradient updates that are unbiased estimates of the total gradient for the
dataset. This is not true on-device, where client updates are biased
representations of the gradient across the entire population~\cite{non_iid_fl}.

Non-IID data adversely affect convergence~\cite{non_iid_conv}, and have been
identified as a fundamental challenge to FL~\cite{adv_op_fl}. Proposals to fit
non-IID data better include optimizers that account for client
drift~\cite{scaffold}, data sharing between client devices~\cite{non_iid_fl},
and adaptive server optimizers with client learning rate
decay~\cite{adaptivefedopt}, among others~\cite{non_iid_class}.

The primary contribution of this paper is to demonstrate that keyword-spotting
models can be trained on large-scale datasets using FL, and can achieve false
accept and false reject rates that rival those of centralized training. Using
simulated federated learning experiments on large-scale datasets consisting
of thousands of speakers and millions of utterances, we address the algorithmic
challenges associated with training on non-IID data, the visibility challenges
associated with labeling on-device data, and the physical constraints that limit
augmentation capabilities on-device.

%% file: model.tex
\section{Model}
\label{sec:model}

\begin{figure}
  \centering
  \includegraphics[width=\columnwidth]{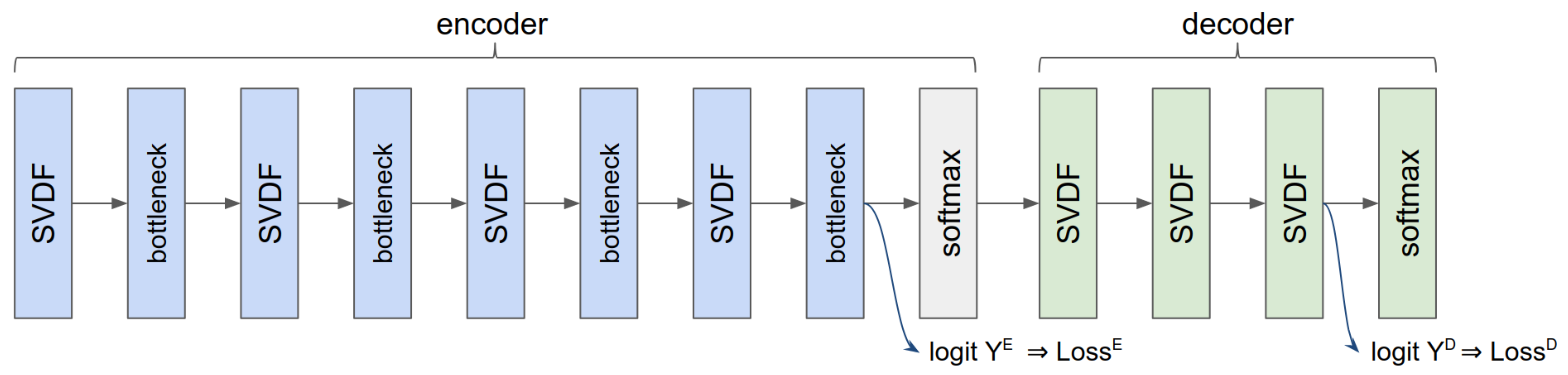}
  \caption{End-to-end topology trained to predict the keyword likelihood
           score~\cite{kws_smp}. The encoder consists of 4 SVDF-plus-bottleneck
           layers, while 3 SVDF layers comprise the decoder.}
  \label{fig:nn_architecture}
\end{figure}

This paper used the end-to-end architecture described in~\cite{kws_smp}.
End-to-end trainable neural architectures have demonstrated state-of-the-art
performance in terms of accuracy as well as lowered resource requirements while
providing a highly optimizable system design~\cite{e2e_kws}. The model consisted
of an encoder-decoder architecture in which both the encoder and decoder made
use of efficiently parameterized SVDF (single value decomposition filter)
layers---originally introduced in~\cite{svdf15}---to approximate fully-connected
layers with low rank approximations. Dense bottleneck layers were used to
further reduce computational costs and keep the model size down at only 320,778
parameters (Figure~\ref{fig:nn_architecture}).

The encoder took spectral domain features $\mathbf{X}_t$ as input and generated
outputs $Y^E$ corresponding to phoneme-like sound units. The decoder
model used the encoder output as input and generated binary output $Y^D$ that
predicted the existence of a keyword. The model was fed with acoustic input
features at each frame (generated every 10ms), and generated prediction labels
at each frame in a streaming manner.

Training such an architecture traditionally required frame-level labels
generated by LVCSR systems~\cite{cnn_kws} to provide accurate timing
information. This approach to label generation is not possible on-device, due to
the large computational resources required to store and run a LVCSR system.
Therefore, while we retained the same architecture, in our experiments we train
the system with just a binary cross entropy loss for keyword presence and did
not present any supervised targets to train the encoder. Recent
work~\cite{kws_smp} suggests a better approach to train the encoder without
LVCSR targets and will be the focus of future work.

%% file: optimization.tex
\section{Federated Optimization}
\label{sec:optimization}

In federated learning~\cite{fedlearn}, a central server sends neural
models to many client devices (such as phones). These clients process local
caches of data in parallel and send updated model weights back to the server.
The server aggregates the updates, produces a new global model, and repeats this
cycle (called a federated training round) until the model converges.

Federated Averaging (\texttt{FedAvg})~\cite{fedlearn} was used as a baseline
optimization algorithm. During each training round, indexed by $t$, a subset of
$K=400$ client devices in the experiment population downloaded a global model,
$w_{t}$, from the server. Each client $k \in K$ had a local data cache
consisting of $n_{k}$ examples. The clients used stochastic gradient descent
(SGD) to train over their local examples and derive an average gradient,
$g_{k}$. For a client learning rate ${\eta}_{c}$, the local client step,
${w}_{t+1}^{k}$, was defined:

\begin{equation}
  {w}_{t+1}^{k} = {w}_{t} - {\eta}_{c} {g}_{k}.
  \label{eq:client_update}
\end{equation}

\noindent
This equation represents a single step of SGD, but client training typically
involved multiple steps of SGD with a batch size of 1. Updated client weights
were sent back to the server, which aggregated them to compute a global model
update:

\begin{equation}
  \Delta_{t} = \sum_{k=1}^{K} \frac{{n}_{k}}{N} ({w}_{t} - {w}_{t+1}^{k}).
  \label{eq:server_delta}
\end{equation}

\noindent
where $N = \sum_{k} {n}_{k}$. For a server learning rate $\eta_{s}$, the updated
global model weights, ${w}_{t+1}$ were computed according to:

\begin{equation}
  {w}_{t+1} = {w}_{t} - \eta_{s} \Delta_{t}.
  \label{eq:server_update}
\end{equation}

\noindent
When phrased in the form of Equation~\ref{eq:server_update}, \texttt{FedAvg}
clearly consists of an \textit{inner optimizer loop} (SGD over gradients on the
clients) and an \textit{outer optimizer loop} (SGD on averaged weight deltas on
the server).

Momentum-based variants of \texttt{FedAvg} were explored as in
Ref.~\cite{flnwp, noniid_mom}, in which Nesterov accelerated
gradients (NAG)~\cite{nesterov} were applied to the server updates. The server
update in Equation~\ref{eq:server_update} was replaced by:

\begin{equation}
  {w}_{t+1} = {w}_{t} - \eta_{s} (\gamma {v}_{t+1} + \Delta_{t}).
  \label{eq:nag_update}
\end{equation}

\noindent
where $\gamma$ is the momentum hyperparameter and
${v}_{t+1} = \gamma {v}_{t} + \Delta_{t}$ is the forward-looking Nesterov
accelerated gradient. The advantages of Nesterov momentum over classical
momentum~\cite{POLYAK19641} have been demonstrated in the central training
setting~\cite{momentumindl}, and were expected to translate to FL.

Finally, adaptive variants of \texttt{FedAvg} were investigated, in which the
server optimizer function was replaced by Adam~\cite{adam}, Yogi~\cite{yogi}, or
LAMB~\cite{lamb}. Prior works shown that adaptive per-coordinate updates can
improve convergence for FL~\cite{fed_kws, adaptivefedopt}. Adaptive methods have
shown particular strength in environments with heavy-tailed stochastic gradient
noise distributions~\cite{adamattention}---a common property of non-IID data in
FL.

Adaptive methods replace the server optimizer loop, shown in
Equation~\ref{eq:server_update}, with an adaptive optimization step. As with
\texttt{FedAvg}, the classical momentum terms of Adam or Yogi can be replaced
with NAG as in NAdam~\cite{nadam}.

%% file: data.tex
\section{Data and Simulations}
\label{sec:data}

Experiments were conducted using simulated federated learning with
vendor-collected datasets. A total of 1.5M utterances (1,650 hours in total)
were recorded by 8,000
English-speaking participants. Each lasted a few seconds, and most contained one
of two spoken keyword phrases. The dataset was divided into a train set (1.3
million utterances) and an eval set (180,000 utterances) with non-overlapping
groups of users. IID and non-IID configurations of the training data were
prepared, while the eval set was always used in an IID configuration.

Utterances were grouped into non-IID simulated clients in three steps. First,
data were clustered according to speaker. The resulting clusters varied
significantly in size: though speakers provided a median of 108 utterances each,
a few speakers were associated with nearly a thousand unique examples. Next,
clusters were further divided on the basis of labels. Individual clusters
contained either positive utterances (which contained the keyword) or negative
utterances (which lacked the keyword) exclusively. It should be noted that
training labels were specified on a per-frame basis, and even positive
utterances contained numerous frames with negative targets. Finally, the
clusters were randomly subdivided to enforce an exponential distribution of
utterances per client ($n_{k}$). The final step was motivated by a desire to
match the training data distribution to the inference distribution, coupled with
observations of power-law feature usage among the general population.

The resulting non-IID training dataset consisted of 160,000 clusters and a
median of 6.5 utterances per cluster. Individual clusters were affiliated with
individual simulated client devices for federated learning.

For IID simulated clients, the data were randomly divided into clusters
consisting of 50 utterances. Each uniformly-sized cluster thus included a mix of
speakers and labels. The IID training set contained 26,000 clusters of 50
utterances each. 3,675 clusters comprised the IID eval set.

%% file: experiments.tex
\section{Experiments}
\label{sec:experiments}

This section describes experiments to address the various constraints of
on-device training. Specifically, we discuss optimization techniques to overcome
the algorithmic challenge of fitting non-IID data, lightweight data augmentation
techniques that run with constrained on-device resources, and teacher-student
training to provide labels given the inability to peek at federated data.

\subsection{Training and evaluation}
\label{sec:train_and_eval}

Model checkpoints generated by the training tasks were periodically saved and
sent to a held-out set of client devices for federated eval tasks. The train and
eval tasks ran over orthogonal datasets, which were constructed following the
description in Section~\ref{sec:data}. Metrics including the frame-level
accuracy and cross-entropy loss were computed on each client and averaged on the
server.

Hyperparameters were tuned and checkpoints were selected based on the criterion
of minimizing eval loss for non-IID data. Eval loss was measured on the IID
dataset described in Section~\ref{sec:data}. Tasks were also trained on IID
datasets in order to compare training under the different data distributions.

Models were evaluated using false accept (FA) and false reject (FR) rates, which
were computed offline on large tests sets consisting of negative utterances and
positive utterances, respectively. The triggering thresholds were tuned to have
FA=0.2\%, approximately. FA and FR are more relevant to quality than loss, since
they directly correspond to inference performance.

\subsection{Optimizers and learning rate schedules}
\label{sec:optimizers}

Optimization techniques were explored for non-IID training. First, the
algorithms described in Section~\ref{sec:optimization} were tuned via grid
searches. For \texttt{FedAvg}, $\eta_{s}=1.0$ and a momentum value of $0.99$ was
found to work best. \texttt{FedAdam} and \texttt{FedYogi} both converged well
with $\beta_{1}=0.9$ and $\beta_{2}=0.999$, though Adam worked with the default
$\epsilon=10^{-8}$ and $\eta_{s}=10^{-3}$ while Yogi worked best with a larger
$\epsilon=10^{-3}$, $\eta_{s}=0.1$, and initial accumulator value of $10^{-6}$.
Experiments were also performed in which Nesterov accelerated gradients were
substituted for classical momenta.

Table~\ref{tab:optimizers} compares non-IID training with each server
optimization algorithm. Exponentially decayed client learning rates were used
with \texttt{FedAdam} and \texttt{FedYogi}, while \texttt{FedAvg} worked better
with constant client learning rates. The adaptive optimizers had a decisive
advantage over \texttt{FedAvg} on FR. Replacing classical momentum with NAG
benefitted \texttt{FedAvg} and \texttt{FedAdam}, but \texttt{FedYogi} with
classical momentum had the lowest FR overall.

\begin{table}[!h]
  \caption{Comparisons of offline false accept and reject rates for various
           optimizers on non-IID data.}
  \centering
  \begin{tabular}{llccc} \toprule
    Optimizer               & FA [\%]  & FR [\%] \\ \hline \midrule
    \texttt{FedAvg}         & 0.21     & 8.76    \\
    \texttt{FedAvg}  + NAG  & 0.21     & 4.09    \\
    \texttt{FedAdam}        & 0.19     & 1.95    \\
    \texttt{FedAdam} + NAG  & 0.21     & 1.68    \\
    \texttt{FedYogi}        & 0.19     & 1.39    \\
    \texttt{FedYogi} + NAG  & 0.20     & 2.11    \\
  \bottomrule
  \end{tabular}
  \label{tab:optimizers}
\end{table}

Theoretical and empirical results indicate that client learning rate (LR) decay
improves convergence on non-IID
data~\cite{non_iid_fl, adaptivefedopt, non_iid_conv}. Fixed client learning
rates (with ${\eta}_{c}=0.02$) were compared with exponentially-decayed client
learning rate schedules, in which ${\eta}_{c}$ was reduced by a constant factor,
${\Gamma}_{\eta, c}$, after every $N_{\Gamma}$ steps. Hyperparameter scans found
that the eval loss was minimized with an initial learning rate
${\eta}_{c, 0}=0.02$, ${\Gamma}_{\eta, c}=0.9$, and $N_{\Gamma}=1000$.

Learning rate comparisons are shown in Table~\ref{tab:learning_rate}, for the
\texttt{FedYogi} optimizer. LR decay significantly improves both IID and non-IID
training. The difference is most pronounced for non-IID training, where the FR
decreases from 2.35\% to 1.39\% given a fixed FA=0.2\%.

\begin{table}[!h]
  \caption{A comparison of client learning rate schedules.}
  \centering
  \begin{tabular}{lcc} \toprule
    LR schedule           & FR (IID) [\%]  & FR (Non-IID) [\%]  \\ \hline \midrule
    \texttt{Constant}     & 2.14           & 2.35 \\
    \texttt{Exponential}  & 1.73           & 1.39 \\
  \bottomrule
  \end{tabular}
  \label{tab:learning_rate}
\end{table}

\subsection{Data Augmentation}
\label{sec:augmentation}

Two common speech data augmentation methods---MTR~\cite{mtr} and
SpecAugment~\cite{specaug}---were tuned for non-IID training.

MTR is an acoustic room simulator that generates noise files which can be
applied to spectrogram inputs. Based on \textit{a priori} distributions, MTR
generates random room sizes and dimensions, speaker and noise source positions,
signal to noise ratios, and reverberation times~\cite{gc_mtr}. The technique is
effective for far-field speech recognition and has been used previously for
keyword spotting~\cite{e2e_kws, kws_smp}.

In the simulation experiments, MTR was used to create up to 100 noised replica
of each clean utterance from vendor data. In order to keep a constant number of
training examples per simulated client, MTR configurations were randomly sampled
every time a given simulated client device was used for training.

Unfortunately, MTR is infeasible for on-device training: users would have to
download additional noise data (for additive noise) and room simulation
configurations (for reverberations). The extra data processing would also
lengthen client training times.

Spectrum augmentation (SpecAugment) is a fast and lightweight alternative for
speech data augmentation. It has been used previously for keyword
spotting~\cite{kws_specaug}, and has been used to achieve state-of-the-art ASR
performance~\cite{large_specaug}. The augmentation policy broadly consists of
three components: (1) \textit{Time Masking}, in which consecutive time frames in
the spectrogram are masked and replaced with Gaussian-distributed noise, (2)
\textit{Frequency Masking}, in which adjacent bins of the spectrogram are
zeroed, and (3) \textit{Time Warping}, in which features are linearly displaced
along the temporal axis.

SpecAugment is an ideal on-device alternative to MTR, as it requires no config
files and minimally increases training time. Tuning in non-IID data simulations
found an optimal configuration of 2 time masks of up to 60 frames along with 2
frequency masks of up to 15 bins. TimeWarp was not used.

\begin{table}[!h]
  \caption{FA and FR comparisons for models trained on IID and non-IID data with
           different data augmentations.}
  \centering
  \begin{tabular}{llcc} \toprule
    Data Augmentation  & Data type  & FA [\%]   & FR [\%] \\ \hline \midrule
    No augmentation    & IID        & 0.17      & 4.20    \\
    No augmentation    & Non-IID    & 0.20      & 3.19    \\
    MTR                & IID        & 0.13      & 6.96    \\
    MTR                & Non-IID    & 0.18      & 6.15    \\
    SpecAugment        & IID        & 0.20      & 1.73    \\
    SpecAugment        & Non-IID    & 0.19      & 1.39    \\
  \bottomrule
  \end{tabular}
  \label{tab:augmentation}
\end{table}

Augmentation strategies for IID and non-IID FL are compared in
Table~\ref{tab:augmentation}. SpecAugment reduced the FR with respect to MTR and
no augmentation on both data distributions. Thus, we can reduce communication
costs and on-device processing time with SpecAugment while also improving
performance.

\subsection{Labeling}
\label{sec:labeling}

High-quality labeling can be difficult to obtain on-device, since peeking at
data is impossible by design in FL, and user feedback signals are unreliable or
infrequent. Given the obstacles to on-device labeling, teacher student training
can be used to adapt a model trained on the server (with manually labeled data)
to the on-device unlabeled data domain~\cite{distillation, domain_ts, mil_hrs}.
Models were trained on both IID and non-IID data with supervised and
teacher-generated labels. For the semi-supervised setting, the teacher model
architecture was identical to the student, but was trained on additional data in
a centralized setting.

\begin{table}[!h]
  \caption{FR comparisons for on-device labeling strategies.}
  \centering
  \begin{tabular}{lcc} \toprule
    Labeling             & FR (IID) [\%]  & FR (Non-IID) [\%]  \\ \hline \midrule
    \texttt{Supervised}  & 1.73           & 1.39 \\
    \texttt{Teacher}     & 2.12           & 2.07 \\
  \bottomrule
  \end{tabular}
  \label{tab:labeling}
\end{table}

Table~\ref{tab:labeling} compares teacher student training with supervised
training. While the FR increases when moving to semi-supervised labels, it is
expected that the matched data available in true on-device data, coupled with
a limited number of samples labeled with user feedback signals, will close the
performance gap.

\subsection{Central training comparison and ablation studies}
\label{sec:ablation}

The previously-discussed techniques were applied and then removed individually
in an ablation study, and the results were compared with a model trained in the
centralized setting on the exact same vendor dataset using $4\times10^{8}$ steps
of asynchronous SGD. This provided a direct comparison of centralized training
and FL on IID and non-IID data.

Two additional techniques were explored to fit non-IID data. Client update
clipping, based on $\Vert {w}_{t+1}^{k} - {w}_{t} \Vert^{2}$, was tuned for
non-IID fitting. Multiple client training epochs were also studied, and have
been shown to improve convergence~\cite{adaptivefedopt}.

The following settings were used for the ablation study FL baseline: data were
augmented with SpecAugment, 10 client epochs were used, the client LR was
decayed, client L2 weight norms were clipped to 20, and the \texttt{FedYogi}
optimizer was used to train with supervised labels.

\begin{figure}
  \centering
  \includegraphics[width=\columnwidth]{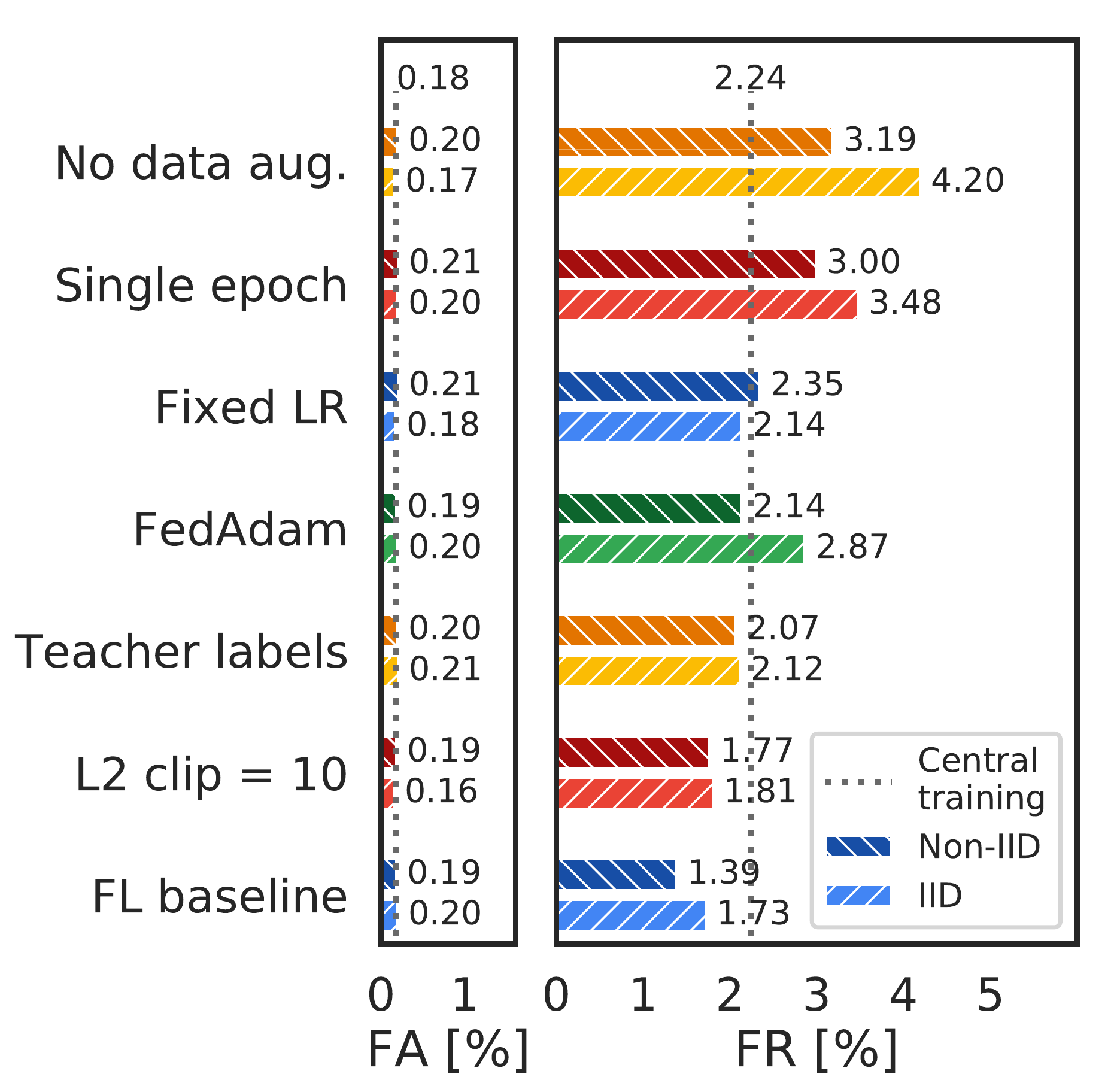}
  \caption{FA and FR for the ablation study, with models trained on IID and
           non-IID data compared with centralized training (dashed line).}
  \label{fig:ablation}
\end{figure}

Ablation results are shown in Figure~\ref{fig:ablation}. The metrics favor
non-IID data because the hyperparameters were tuned to minimize non-IID eval
loss. Had the hyperparameters been tuned for IID data, the IID FR would be lower
than the FR tuned for non-IID data.

FL achieves comparable FR performance with the centrally-trained model. In
absolute terms, SpecAugment and multiple client epochs provided the largest
contributions to both IID and non-IID performance. Interestingly, decayed client
learning rates were more important to non-IID training than IID training. And
contrary to prior studies~\cite{non_iid_fl}, we found that additional client
training epochs benefitted non-IID training.

%% file: conclusions.tex
\section{Conclusions}
\label{sec:conclusions}

Empirical studies were conducted to train a keyword-spotting model using FL on
non-IID data. Adaptive server optimizers like \texttt{FedYogi} helped train a
model with a lower false reject rate in fewer training rounds. We also
demonstrated the necessity and utility of replacing MTR with SpecAugment for
on-device training. Ablation studies revealed the importance of multiple
client epochs and reduced client clipping. And we provided strong empirical
evidence in favor of client learning rate decay for training with non-IID data.
Finally, we overcome the visiblity limitations of on-device training by
demonstrating that, in the absence of high-quality on-device labels,
teacher-student training can achieve comparable performance.

%% file: acknowledgements.tex
\section{Acknowledgements}
\label{sec:acknowledgements}

The authors would like to thank Google Research colleagues for providing the FL
framework, Manzil Zaheer for his optimizer expertise, and Daniel Park for
SpecAugment discussions.